\documentclass[preprint,12pt]{elsarticle}
\usepackage{amsmath}
\usepackage{amssymb}

\usepackage{xcolor}
\usepackage{textgreek}
\usepackage[utf8]{inputenc}
\usepackage[english]{babel}
\usepackage{hyperref}
\usepackage{color,colortbl}
\usepackage{tensind}
\tensordelimiter{?}
\DeclareGraphicsExtensions{.bmp,.png,.jpg,.pdf}
\usepackage{verbatim}
\usepackage[normalem]{ulem}
\usepackage{orcidlink}
\usepackage{soul}
\journal{Physics Letters B}
\begin{document}
\title{Electromagnetic Radiation from Binary Stars Mediated by Ultralight Scalar}

\author[inst1]{Ya-Ze Cheng\orcidlink{0009-0005-7458-5640}}
\author[inst2]{Wen-Hao Wu\orcidlink{0009-0007-1676-0652}}
\author[inst3]{Yan Cao\orcidlink{0009-0002-3959-5059}}
\affiliation[inst1]{organization={School of Astronomy and Space Sciences, University of Chinese Academy of Sciences (UCAS)}, city={Beijing}, postcode={100049},country={China}}
\affiliation[inst2]{organization={School of Physics, Peking University}, city={Beijing},postcode={100871},country={China}}
\affiliation[inst3]{organization={School of Physics, Nanjing University}, city={Nanjing}, postcode={210093},country={China}}

\begin{abstract}
We present the electromagnetic (EM) dipole radiation flux from an eccentric Keplerian binary endowed with scalar charges, in the presence of scalar-photon coupling $\phi A_\mu A^\mu$ or $\phi F_{\mu\nu}F^{\mu\nu}$. The scalar radiation is suppressed for orbital frequency below the scalar mass, while the scalar-mediated indirect EM radiation survives. We examine the constraints imposed on the scalar-photon and scalar-charge couplings by the current observational data of pulsar binaries, in case that the scalar charge is given by the muon number. The general extensions of the calculation to the quadrupole order and hyperbolic orbit are also discussed.
\end{abstract}
\maketitle
\section{Introduction}
Observations have shown that for several pulsar (PSR) binaries, the measured orbital period decay matches the result predicted by vacuum general relativity with remarkable precision. If the binary is sufficiently isolated from its environment, deviations from this result could then reveal the possible modifications on the binary's intrinsic (conservative or radiative) dynamics, such as the radiation of hidden ultralight bosonic particles due to their non-gravitational couplings with the binary's microscopic constituents. Such ultralight bosons are predicted by a wide class of theories and are good candidates of dark matter \cite{Peccei.Quinn, Wilczek.1978, Preskill:1982cy, Abbott:1982af, Dine:1982ah,Holdom:1985ag, Svrcek:2006yi, Arvanitaki_2010, Hui:2016ltb, Hui:2021tkt, Ferreira_2021, 2168507}. So long as the wavelength of the radiation\footnote{It is given by $\lambda=\frac{2\pi}{|\mathbf{k}|}=\frac{2\pi}{\sqrt{\omega^2-m^2}}$, where $m$ is the particle mass. Since $\omega\sim\Omega=2\pi/T$, in the massless limit $\lambda\sim T$. For $\Omega\sim \sqrt{M/r^3}$ and $r\gg M$, this condition is met since $T\gg r$ and the binary separation $r$ is much larger than the size of each body.} and the boson's Compton wavelength are much larger than the size of an individual star, the star can effectively be treated as a point charge. The dipole formula of massive scalar and vector field radiation from a charged eccentric Keplerian binary was derived in \cite{PhysRevD.49.6892}, and the effects of scalar and vector charges on the  binary's orbital dynamics and the resultant astrophysical constraints have been extensively studied, see \cite{Mohanty:1994yi,Hook:2017psm,Huang:2018pbu,KumarPoddar:2019ceq,KumarPoddar:2019jxe,PhysRevD.102.023005,Seymour:2020yle,Hou:2021suj,Gupta:2022spq,Bhattacharyya:2023kbh,Diedrichs:2023foj,Bai:2024pki,Liu:2025zuz} for an incomplete list, and see \cite{Cardoso:2018zhm,Poddar:2021yjd,PhysRevD.107.044041} for discussions of massive tensor radiation.

The radiation flux of massive bosonic fields from a charged eccentric binary (henceforth referred as the \textit{direct} radiation) is suppressed if the orbital frequency is below the boson mass, due to the sharp decrease of Bessel function $|J_n(ne)|$ with the harmonic number $n$. The higher-order process mediated by the boson, from its couplings to even lighter particles, can become significant in this regime, generating additional \textit{indirect} radiation. In \cite{Gavrilova:2023lxu} a coupling between the scalar or vector boson and an ultralight Dirac fermion was considered, although it was shown that similar processes in the standard model (SM) are negligible, a sizable indirect radiation could nonetheless arise due to physics beyond SM (BSM).

In this paper, we explore another possibility, namely a coupling $\phi A_\mu A^\mu$ or $\phi F_{\mu\nu}F^{\mu\nu}$ between a real massive scalar $\phi$ sourced by the binary and a real massless vector $A_\mu$, we derive the dipole radiation flux of the vector field from an eccentric Keplerian binary endowed with scalar charges\footnote{The possibilities of indirect photon radiation from a massive vector mediator sourced by the binary is briefly discussed in \ref{appendix_2}.}. The features of the proposed indirect radiation are analyzed and further illustrated by a concrete scenario in which the scalar charge is given by the muon content of the star, for this model we place simultaneous constraints on the coupling strength using the observational data of two pulsar binaries. In case of the vector particle being the SM photon, the indirect EM radiation itself might also enable such couplings to be probed or constrained.

This paper is organized as follows: In Sec.~\ref{sec2}, we present the energy flux of the direct and indirect radiation due to the ultralight scalar in the relevant models, and analyze their main features. In Sec.~\ref{sec3}, the observational constraints on a concrete scenario are examined. Sec.~\ref{sec4} is a brief summary. In \ref{appendix_1}, \ref{appendix_2}, \ref{appendix_3}, \ref{appendix_4}, we discuss the extensions of the calculation to the quadrupole order, angular momentum radiation and hyperbolic orbit, in particular we derive the quadrupole energy flux and dipole angular momentum flux of massive scalar and vector radiation from a charged eccentric Keplerian binary, which can be useful for a general investigation on the adiabatic orbital evolution. Throughout this paper, we use the flat spacetime metric $\eta_{\mu\nu}=\text{diag}(1,-1,-1,-1)$ and natural units $\hbar=c=G=1$, also we use the 4-momentum notation $k=\{k^\mu\}_\mu=(\omega,\mathbf{k})$, $k_1\cdot k_2\equiv (k_1)_\mu (k_2)^\mu$ with $d^3k\equiv dk_xdk_ydk_z$ and $\int \frac{d^3k}{(2\pi)^3}=\int_\mathbf{k}$, the element of solid angle is denoted by $d\Omega_{\mathbf{k}}$. The complex conjugate of a quantity $X$ is denoted by $\bar X$ and its mass dimension denoted by $[X]$, e.g., $[\phi]=[A_\mu]=[h_{\mu\nu}]=[\Omega]=1$.

\section{Binary Radiation Power}\label{sec2}
Consider a Keplerian binary with orbital period $T=2\pi/\Omega$, semi-major axis $a$, eccentricity $e$, mass $M_{1,2}$, time-independent scalar charge $N_{1,2}$ and trajectory $X^\mu_{1,2}=(t,\mathbf{X}_{1,2})$. Also we introduce the reduced mass $\mu\equiv M_1M_2/{M_\text{tot}}$ (with $M_\text{tot}\equiv M_1+M_2$) and the charge-to-mass ratio difference $D\equiv {N_1}/{M_1}-{N_2}/{M_2}$. The binary is assumed to be non-relativistic, hence $v_I^\mu\equiv dX_I^\mu/dt\approx(1,\dot{\mathbf{X}}_I)$. In the flat spacetime approximation, the system can be written as
\begin{footnotesize}
\begin{equation}\label{system}
    \mathcal{L}=\mathcal{L}_\text{binary}+\mathcal{L}_h-\frac{\sqrt{32\pi}}{2}T^{\mu\nu}h_{\mu\nu}+\frac{1}{2}\partial^\mu\phi\,\partial_\mu\phi
-\frac{1}{2}m^2\phi^2+gn\phi+\mathcal{L}_\text{int}-\frac{1}{4}F^{\mu\nu}F_{\mu\nu},
\end{equation}
\end{footnotesize}
where $\mathcal{L}_h$ is the kinetic term of graviton $h_{\mu\nu}\equiv (g_{\mu\nu}-\eta_{\mu\nu})/\sqrt{32\pi}$, $A_\mu$ is a real massless photon ($\gamma$) with $F_{\mu\nu}=\partial_\mu A_\nu-\partial_\nu A_\mu$, and $\phi$ a real massive scalar boson with coupling $g$ to the scalar charge density\footnote{This is the leading-order description, hence the effects of potential modes on the radiation are not captured. Those subleading relativistic corrections can be taken into account in a higher-order perturbative EFT treatment by computing the corrections to the effective source term in a flat spacetime background, which we do not consider in this paper.} $n(t,\mathbf{x})=\sum_{I=1,2} N_I\,\delta^3(\mathbf{x}-\mathbf{X}_I(t))$, we also include a scalar-photon coupling denoted by $\mathcal{L}_\text{int}$.

We focus on the case of an elliptical orbit, which can be parameterized by the eccentric anomaly $\xi$ as
\begin{equation}
X(t)=a(\cos\xi-e),\quad Y(t)=a\sqrt{1-e^2}\sin\xi,\quad Z(t)=0,\quad\Omega t=\xi-e\sin\xi,
\end{equation}
where $\mathbf{X}\equiv \mathbf{X}_1-\mathbf{X}_2=(X,Y,Z)$ is the relative position of the two bodies written in a Cartesian coordinate frame centered at the binary's mass center such that $\mathbf{X}_2=-(M_1/M_\text{tot})\,\mathbf{X}$ and $\mathbf{X}_1=(M_2/M_\text{tot})\,\mathbf{X}$, with $X$-axis parallel to the major axis and $Z$-axis normal to the orbital plane. For the radiation process from a temporally periodic classical source with $N$-body final state, the time-averaged energy radiation flux is given by its amplitude $\mathcal{M}$ \cite{KumarPoddar:2019ceq,Gavrilova:2023lxu}, which is
\begin{equation}\label{radiation_power_formula}
P=\frac{\Delta E}{T}=\sum_{n=n_\text{min}}^{\infty}P_n=\sum_{n=n_\text{min}}^{\infty}\frac{1}{S}\int\Omega_n\prod_{i=1}^Nd\Pi_i\,2\pi\,\delta\left(\sum_i\omega^{(i)}-\Omega_n\right)|\mathcal{M}_n|^2,
\end{equation}
where $d\Pi_i\equiv\frac{d^3k^{(i)}}{(2\pi)^32\omega^{(i)}}$, $S$ is the symmetry factor of this process. The source has been decomposed into a Fourier series with oscillation frequency $\Omega_n\equiv n\Omega$, and $n_\text{min}$ is the minimal value of $n$ for which $P_n$ is nonzero. In the present case there are three main contributions to the radiation:\footnote{Similar to scalar-mediated process, there is graviton-mediated particle production from the minimal graviton-matter coupling and the graviton self-interaction, but it is clearly negligible. The radiation fields themselves can also generate secondary graviton and photon radiation, although this is not a part of the binary's dissipative dynamics.}
\begin{equation}
P=P^{(0)}+P^{\text{(I)}}+P^{\text{(II)}},
\end{equation}
which are the gravitational radiation, the direct scalar radiation and the indirect scalar-mediated photon radiation, respectively.

From Eq.~\eqref{system}, the amplitude of gravitational radiation is given by $i\mathcal{M}_n=-i\frac{\sqrt{32\pi}}{2}T^{\mu\nu}(\Omega_n,\mathbf{k})\,\bar\epsilon_{\mu\nu}^{(\lambda)}(\mathbf{k})$, where $\epsilon_{\mu\nu}^{(\lambda)}(\mathbf{k})$ is the normalized polarization tensor of graviton with $\epsilon_{\mu\nu}{\bar \epsilon}^{\mu\nu}=1$ (see also \cite{Mohanty:1994yi,Poddar:2021yjd}), and $T^{\mu\nu}(\Omega_n,\mathbf{k})$ is the Fourier transform of the energy-momentum tensor (EMT). Using the approximation $e^{i\mathbf{k}\cdot{\mathbf{x}}}\approx 1$ in evaluating the Fourier transform of $T^{ij}(t,\mathbf{x})$\footnote{In the case of binary, it is not merely given by the matter EMT: $T^{ij}(t,\mathbf{x})\approx\sum_{I=1,2}M_I \dot X_I^i \dot X_I^j\delta^3(\mathbf{x}-\mathbf{X}_I)$ and could instead be derived from $T^{00}(t,\mathbf{x})\approx \sum_{I=1,2} M_I\delta^3(\mathbf{x}-\mathbf{X}_I)$ via $\partial_\mu T^{\mu\nu}\approx\nabla_\mu T^{\mu\nu}=0$, the result in the limit $|\mathbf{k}\cdot\mathbf{x}|\ll 1$ is $T^{ij}(\omega,\mathbf{k})=\int d^3x\,e^{-i\mathbf{k}\cdot\mathbf{x}}T^{ij}(\omega,\mathbf{x})\approx-\frac{1}{2}\omega^2\int d^3x\,T_{00}(\omega,\mathbf{x})\,x^ix^j$. This is equivalent to including the EMT of the Newtonian potential, such that $\int d^3x\,T^{ij}(t,\mathbf{x})\approx\mu\left(\dot X^i \dot X^j-\frac{M_\text{tot}}{r^3}X^iX^j\right)$.}, we obtain the leading-order gravitational radiation power (Peters-Matthews formula):
\begin{equation}\label{P_0}
P^{(0)}=\sum_{n=1}^\infty P^{(0)}_n=\frac{32}{5}a^4\mu^2\Omega^6
\frac{37 e^4+292 e^2+96}{96 \left(1-e^2\right)^{7/2}}.
\end{equation}

The amplitude of scalar radiation is given by $i\mathcal{M}_n=ign(\Omega_n,\mathbf{k})$. In the dipole approximation,
\begin{equation}\label{dipole}
\begin{aligned}
    n(\Omega_n,\mathbf{k})
&=
\frac1T\int_0^Tdt\int d^3x\, e^{-i\mathbf{k}\cdot\mathbf{x}+i\Omega_n t}n(t,\mathbf{x})
\\
&\approx
\frac1T\int_0^Tdt\sum_{I=1,2}N_I\,[-i\mathbf{k}\cdot\mathbf{X}_I(t)]\,e^{i\Omega_nt}=
a\mu D \,\mathbf{j}_n\cdot\mathbf{k}
,
\end{aligned}
\end{equation}
where
\begin{equation}\label{j_n}
\mathbf{j}_n\equiv
\frac{1}{n}
\left(\begin{matrix}
-i J'_n
\\
\frac{(1-e^2)^{1/2}}{e}J_n
\\
0
\end{matrix}\right),
\quad
J_n\equiv J_n(ne),
\quad
J'_n\equiv \left.\frac{dJ_n(z)}{dz}\right|_{z=ne},
\end{equation}
here $J_n(z)$ is the Bessel function of the first kind. In the large-$n$ limit (see 9.3.2 of \cite{abramowitz1968handbook}), 
\begin{equation}
J_n(ne)\to \frac{\exp\left[\left(\sqrt{1-e^2}-\text{arccosh}\,e^{-1}\right)n\right]}{\sqrt{2\pi \sqrt{1-e^2}\,n}}.
\end{equation}
Since $\mathbf{j}_n$ in independent of $\mathbf{k}$, the integration in Eq.~\eqref{radiation_power_formula} can be simplified by the fact $\int d\Omega_\mathbf{k}\,k_ik_j=\frac{4\pi}{3}|\mathbf{k}|^2\delta_{ij}$, the final result for the radiation power is (see also \cite{PhysRevD.49.6892,Mohanty:1994yi})
\begin{align}\label{P_I}
P^\text{(I)}&=\sum_{n=\lceil n_0 \rceil}^\infty P_n^\text{(I)},
\\
P_n^\text{(I)}&=\frac{1}{6\pi}g^2a^2{\mu}^2D^2\Omega^4n^2\left[(J_n')^2+\frac{1-e^2}{e^2}(J_n)^2\right]\left(1-\frac{n_0^2}{n^2}\right)^{3/2},
\end{align}
with $n_0\equiv m/\Omega$ and $\lceil x\rceil$ denotes the smallest integer larger than or equal to $x$. A closed-form result can only be obtained for circular orbit or in the massless limit:
\begin{align}
P^\text{(I)}(e=0)&=\frac{1}{12\pi}g^2 a^2\mu^2D^2\Omega ^4\left(1-n_0^2\right)^{3/2}
,
\\
P^\text{(I)}(m=0)&=\frac{1}{12\pi}g^2a^2\mu^2D^2\Omega^4\frac{(1+e^2/2)}{(1-e^2)^{5/2}}.\label{P_I_massless}
\end{align}

For the indirect radiation due to the scalar-photon coupling, we consider two models below.

\subsection{Model I}\label{P_II_model_1}
\begin{figure}[hbt]
	\centering
	\includegraphics[width=0.6\textwidth]{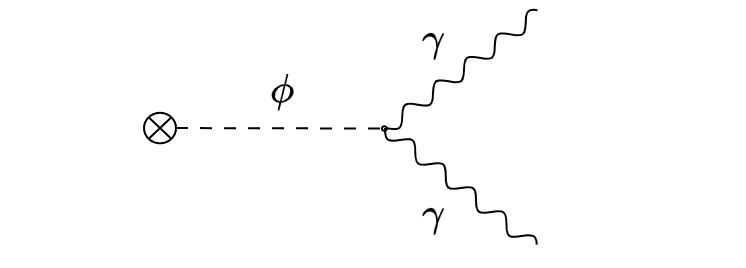}
	\caption{The scalar-mediated photon radiation channel considered in this paper. It is not a radiation process in the classical sense (classically the scalar field cannot be a source for the photon in the absence of a background EM field), rather it originates from the (quantum-mechanical) spontaneous decay of $\phi$ sourced by the binary due to its orbital motion.}\label{diagram}
\end{figure}

The first model is given by a coupling between $\phi$ and the lowest-dimensional photon operator:
\begin{equation}\label{scalar}
\mathcal{L}_\text{int}=\frac{1}{2}g'\phi A^\mu A_\mu,
\end{equation}
with the mass dimension of the coupling constant $g'$ being $[g']=1$ (in contrast to $[g]=0$). The resultant radiation process is depicted in Fig.~\ref{diagram}. Using a Breit-Wigner propagator \cite{Tait2009TASILO}, the matrix element of this process is
\begin{equation}
    i \mathcal{M}_n = ign(\Omega_n,\mathbf{k})\,\frac{i}{k^2-m^2 + im\Gamma_{\phi}}\,i g'\eta^{\mu\nu}\bar\epsilon_{\mu}^{(\lambda_1)} (\mathbf{k}_1)\,\bar\epsilon_{\nu}^{(\lambda_2)}(\mathbf{k}_2),
\end{equation} 
with $k=k_1+k_2$ and $\Gamma_\phi=\frac{(g')^2}{16\pi m}$ the decay width of the $\phi\to \gamma\gamma$ process.

Using the explicit polarization sum of massless photon (see, e.g., Sec. 6.4 of \cite{Greiner:1996zu}):
\begin{equation}
\sum_{\lambda}\epsilon_\mu^{(\lambda)}(\mathbf{k})\,\bar \epsilon_\nu^{(\lambda)}(\mathbf{k})=-\eta_{\mu\nu}-\frac{k_\mu k_\nu}{\omega^2}+\frac{k_\mu n_\nu+k_\nu n_\mu}{\omega},
\end{equation}
where the 4-vector $n$ satisfies $k\cdot n=\omega$, we obtain\footnote{If the photon is massive, the result would be $2+\frac{(k_1\cdot k_2)^2}{(k_1\cdot k_1)(k_2\cdot k_2)}$.}
\begin{equation}
\sum_{\lambda_1,\lambda_2}\left[\epsilon^{(\lambda_1)}(\mathbf{k}_1)\cdot \bar\epsilon^{(\lambda_2)}(\mathbf{k}_2)\right]\left[\bar\epsilon^{(\lambda_1)}(\mathbf{k}_1)\cdot \epsilon^{(\lambda_2)}(\mathbf{k}_2)\right]=2,
\end{equation}
and
\begin{equation}
    \sum_{\lambda_1,\lambda_2} |\mathcal{M}_n|^2 = \frac{2g^2 (g')^2 |n(\Omega_n,\mathbf{k})|^2}{(k^2-m^2)^2+ m^2 \Gamma_{\phi}^2},
\end{equation}
hence (in this integral $|\mathbf{j}_n\cdot\mathbf{k}|^2$ can also be effectively replaced by $|\mathbf{j}_n|^2|\mathbf{k}|^2/3$)
\begin{equation}\label{model_I}
\begin{split}
    P_n^{\text{(II)}}&=\,\frac12 \int d\Pi_1 d\Pi_2 \,2\pi\,\delta(\Omega_n-\omega_1-\omega_2)\, 
    \Omega_n \left[
    \frac{4g^2 (g')^2a^2\mu^2 D^2|\mathbf{j}_n\cdot\mathbf{k}|^2}{(k^2-m^2)^2+ m^2 \Gamma_{\phi}^2}
    \right]
    \\
    &=\,\frac{1}{96{\pi}^3}g^2(g')^2a^2{\mu}^2D^2\Omega^2n^{-1}\left[(J_n')^2+\frac{1-e^2}{e^2}(J_n)^2\right]\int_{0}^{n}dx\,F(x),
\end{split}
\end{equation}
with $x= \omega_1/\Omega$, $\omega_2/\Omega=n-x$,
\begin{equation}\label{auxiliary}
    A\equiv \frac{n_0^2}{2x(n-x)}, \quad B\equiv \frac{n_0^2n_{\Gamma}^2}{4x^2(n-x)^2}, \quad C\equiv \frac{x^2+(n-x)^2}{2x(n-x)},
    \quad
\end{equation}
where $n_0\equiv \frac{m}{\Omega},n_{\Gamma}\equiv \frac{\Gamma_\phi}{\Omega},$ and
\begin{equation}\label{F_model_I}
\begin{aligned}
    F(x)&\equiv \int_0^{\pi} d\gamma\, \sin \gamma\,\frac{C+\cos\gamma}{(1-A-\cos\gamma)^2+B}
    \\
    &=\frac{1}{2}\ln{\frac{A^2+B}{(A-2)^2+B}}+\frac{1-A+C}{\sqrt{B}}\left[\arctan\left({\frac{A}{\sqrt{B}}}\right)-\arctan\left({\frac{A-2}{\sqrt{B}}}\right)\right].
\end{aligned}
\end{equation}
In the case of circular orbit, $\lim_{e\to 0}\left[(J_n')^2+\frac{1-e^2}{e^2}(J_n)^2\right]=\frac{1}{2}\delta_{n,1}$, the radiation power is given by
\begin{equation}
P^{\text{(II)}}=\frac{1}{192{\pi}^3}g^2(g')^2a^2{\mu}^2D^2\Omega^2\int_{0}^{1}dx\,F(x).
\end{equation}

We note that the above results also apply for the indirect radiation in a model $\mathcal{L}_\text{int}=\sqrt{2}g'\phi\varphi^2$, replacing the vector $A_\mu$ with a massless real scalar $\varphi$.

\subsection{Model II}\label{P_II_model_2}
For the second model, we consider a dilatonic coupling:
\begin{equation}
\mathcal{L}_\text{int}=\frac{1}{4}g'\phi F^{\mu\nu}F_{\mu\nu},
\end{equation}
with $[g']=-1$. Similar calculation gives the indirect radiation power:\footnote{
The result is same for a (pseudoscalar) axionic coupling $\mathcal{L}_\text{int}=\frac{1}{2}g'\phi \tilde F^{\mu\nu}F_{\mu\nu}$ with $\tilde{F}^{a b}\equiv\frac{1}{2} \epsilon^{a b c d} F_{c d}$.
}
\begin{equation}\label{model_II}
P^\text{(II)}_n=\frac{1}{48{\pi}^{3}}g^{2}(g')^2a^{2}\mu^{2}D^{2}\Omega^{6}n^{-1}\left[(J_n')^2+\frac{1-e^2}{e^2}(J_n)^2\right]\int_{0}^{n}dx\,F(x),
\end{equation}
with
\begin{equation}\label{F_model_II}
\begin{aligned}
 F(x)\equiv &\,x^2(n-x)^2 \int_0^{\pi} d\gamma \,\sin \gamma \,  \frac{(1-\cos \gamma)^2(C+\cos \gamma)}{(1-A-\cos \gamma)^2+B}
 \\
 =&\,x^2(n-x)^2\bigg\{F_0(x)+F_1(x)\left[\arctan\left(\frac{A}{\sqrt{B}}\right)-\arctan\left(\frac{-2+A}{\sqrt{B}}\right) \right]
 \\
 &\,+F_2(x)\,\text{arctanh}\left[\frac{2(A-1)}{2+B+A(A-2)}\right]\bigg\},
\end{aligned}
\end{equation}
with the decay width $\Gamma_{\phi} = \frac{(g')^2m^3}{32\pi}$ and
\begin{align}
F_0(x) &\equiv 2\left(C-2A\right),
\\
F_1(x) &\equiv \frac{1}{\sqrt{B}}\left[-A^3+A^2(C+1)+3AB-B(C+1)\right],
\\
F_2(x) &\equiv 3A^2-2A(C+1)-B,
\end{align}
where the definitions of $A,B,C,n_0,n_\Gamma$ are identical with Eq.~\eqref{auxiliary}. Note that in the limit $g'\to 0$, the summation has to be restricted to $1\le n\le n_0$, so there would be no indirect radiation if $\Omega>m$.

\begin{table}[hbt]
\center
\renewcommand{\arraystretch}{2}
\resizebox{\textwidth}{!}{
        \begin{tabular}{|c|c|c|}
        \hline
         \textbf{Limits} & \textbf{Model I (Sec.~\ref{P_II_model_1})} & \textbf{Model II (Sec.~\ref{P_II_model_2})}
         \\
         \hline
         $\Omega\to 0$ &
         $\int_{0}^{n}dx\,F=\frac{512 \pi^2 n^5  \Omega^4}{5{(g')}^4 +256 n_0^4\Omega^4 \pi^2}$
         & $\int_{0}^{n}dx\,F=
    \frac{4096 \pi ^2 n^9 \Omega ^4}{105m^4 \left[3 (g')^4 m^4+3072 \pi ^2\right]}$
    \\ 
    \hline
         $\Omega\to \infty$ &
         $F=
    \frac{16{\pi}n^2{\Omega}^2}{(g')^2}\left\{\text{arctan}\,\frac{16{\pi}n^2\Omega^2}{(g')^2}-\text{arctan}\,\frac{16{\pi}[n^2-4\pi x(n-x)]\Omega^2}{(g')^2}\right\}
    $
    &  
    $\int_{0}^{n}dx\,F=\frac{n^5}{10}$
    \\
    \hline
         $g'\to 0$ &
         $F=\ln\frac{{n_0}^2}{4 x (x-n)+{n_0}^2}+\frac{4 x ({n_0}-n) (n+{n_0}) (x-n)}{4 {n_0}^2 x (x-n)+{n_0}^4}$
         & 
         $F=\frac{2 n^2 \Omega ^2-3 m^2}{4 \Omega ^4m^{-2}}\ln \frac{m^2+4 x \Omega ^2 (x-n)}{m^2}+\frac{-3 m^4+2 m^2 \Omega ^2 \left(n^2+3 n x-3 x^2\right)+4 x \Omega ^4 (x-n) \left(n^2-2 n x+2 x^2\right)}{m^2 \Omega ^2x^{-1} (n-x)^{-1}-4 \Omega ^4}$ 
    \\
    \hline
         $g'\to \infty$
         &
         $\int_{0}^{n}dx\,F=\frac{512\pi^2 n^5  \Omega^4}{5 {(g')}^4}$
         &
         $\int_{0}^{n}dx\,F=\frac{4096 \pi ^2 n^9 \Omega^4}{315 (g')^4 m^8}$
    \\
    \hline
         $m\to 0$
         &
         $F=\frac{1}{2}\ln \frac{(g')^4}{(g')^4+4096 \pi ^2 x^2 \Omega ^4 (n-x)^2}+16 \pi  n^2 \frac{\Omega ^2}{ (g')^2}\arctan\frac{64 \pi  x \Omega ^2 (n-x)}{(g')^2}$
         &
         $ \int_{0}^{n}dx\,F=\frac{8 \pi ^2 n^3}{(g')^2\Omega^2}+\frac{n^5}{10} $
    \\
    \hline
         $m\to \infty$
         &
         $ \int_{0}^{n}dx\,F=\frac{2 n^6}{5 n_0^4}$
         &
         $ \int_{0}^{n}dx\,F =\frac{16384 \pi ^2 n^5 \Omega^4}{45 (g')^4 m^8}$
         \\         
         \hline
        \end{tabular}
        }
    \caption{The asymptotic limits of $P^\text{(II)}(g',m,\Omega)$.}
    \label{table1}
\end{table}

\subsection{Asymptotic Limits}
Since $n_0=m/\Omega$, $n_\Gamma\propto (g')^2/\Omega$, and $F(x)$ is a function of $(n_0,n_\Gamma)$, the $(g',\Omega,e)$-dependence of the radiation power is fully captured by the following dimensionless characteristic functions:
\begin{align}
    D_{g'}(n_{\Gamma},n_0,e)&=\sum_{n=1}^{\infty}\frac{n_{\Gamma}}{n}\left[(J_n')^2+\frac{1-e^2}{e^2}{J_n}^2\right]\int_{0}^{n}dx\, F(x),
\\
    D_{\Omega}(n_{\Gamma},n_\Gamma/n_0,e)&=\sum_{n=1}^{\infty}\frac{{n_{0}}^{-s}}{n}\left[(J_n')^2+\frac{1-e^2}{e^2}{J_n}^2\right]\int_{0}^{n}dx\, F(x),
\end{align}
with the parameter choice $s=2,6$ for model I, II, respectively; the direct scalar radiation corresponds effectively to $\int_0^ndx\,F(x)\propto n^3(1-n_0^2/n^2)^{3/2}$ with $s=4$, $n_\Gamma=0$ and $n\ge n_0$.

\begin{figure}[hbt]
	\centering
	\includegraphics[width=0.47\textwidth]{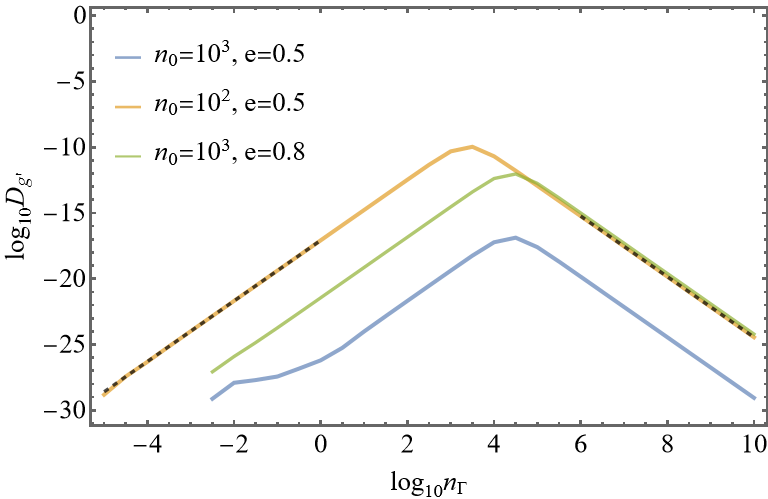}
\quad
    \includegraphics[width=0.47\textwidth]{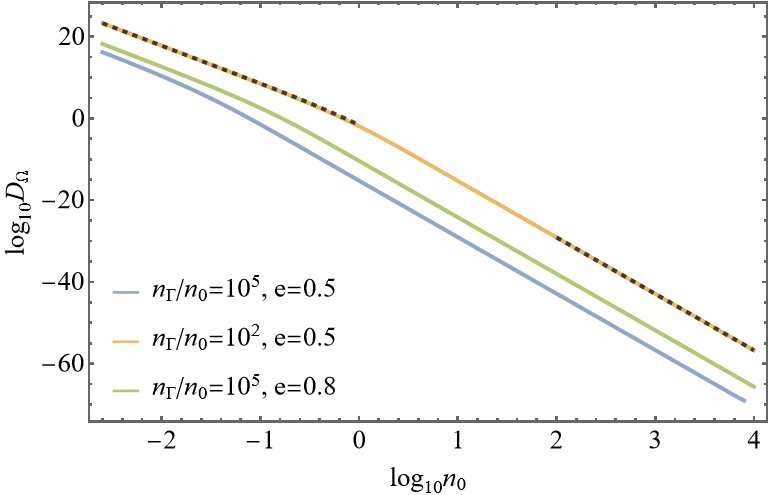}
	\caption{The asymptotic limits of $P^\text{(II)}$ in model I. Left: $P^\text{(II)}(g')\propto D_{g'}(n_\Gamma)$ for given $n_0$ and $e$. Right: $P^\text{(II)}(\Omega)\propto D_\Omega (n_0)$ for given $n_\Gamma/n_0\propto (g')^2$ and $e$.}\label{fig:model_I}
\end{figure}

\begin{figure}[hbt]
	\centering
	\includegraphics[width=0.47\textwidth]{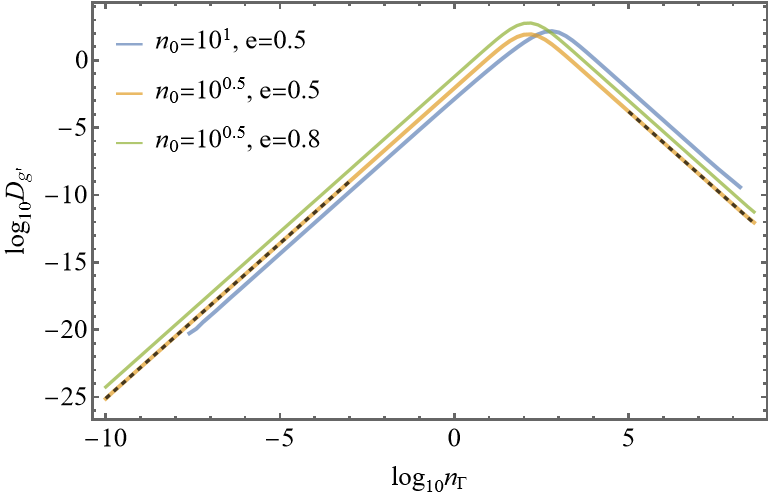}
\quad
    \includegraphics[width=0.471\textwidth]{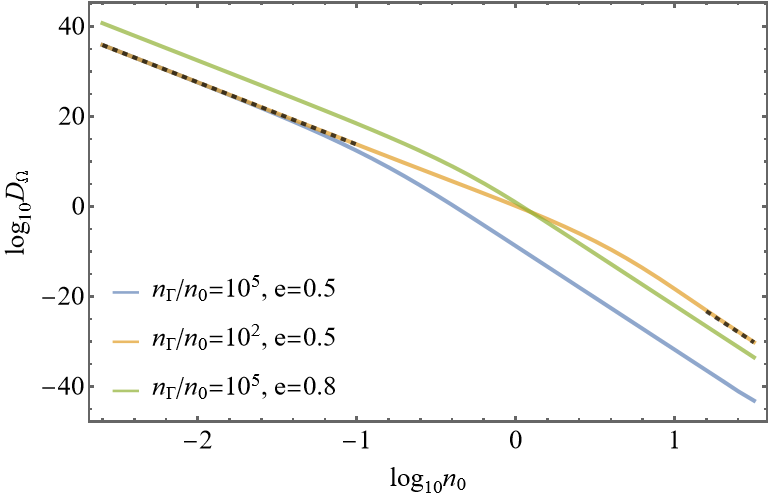}
	\caption{The asymptotic limits of $P^\text{(II)}$ in model II. Left: $P^\text{(II)}(g')\propto D_{g'}(n_\Gamma)$ for given $n_0$ and $e$. Right: $P^\text{(II)}(\Omega)\propto D_\Omega (n_0)$ for given $n_\Gamma/n_0\propto (g')^2$ and $e$.}\label{fig:model_II}
\end{figure}

\begin{figure}[hbt]
	\centering
	\includegraphics[width=0.47\textwidth]{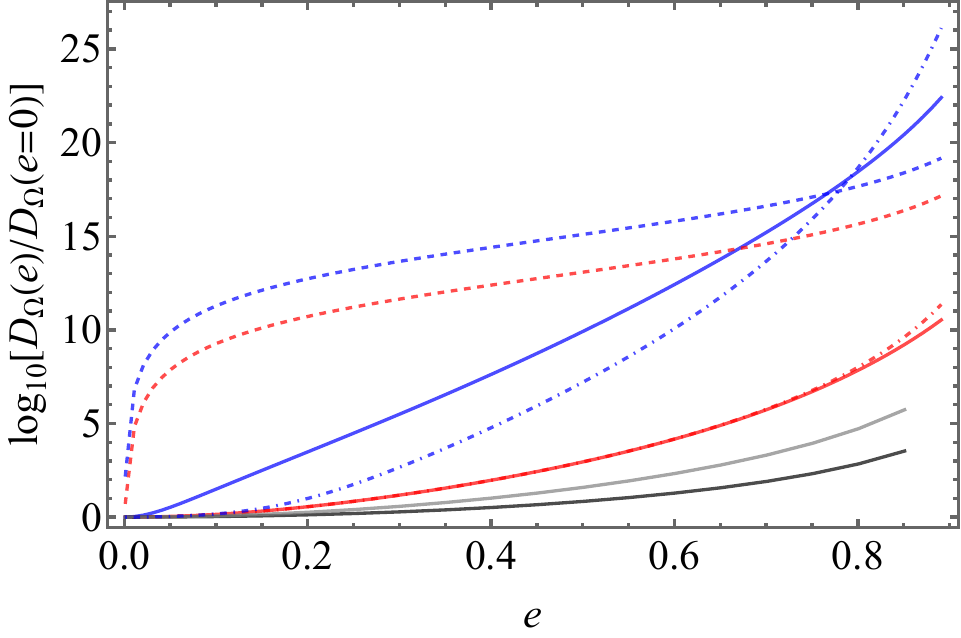}
	\caption{Enhancement of the radiation power with orbital eccentricity as measured by the ratio $P(e)/P(e=0)=D_\Omega(e)/D_\Omega(e=0)$,with $(n_\Gamma/n_0,n_0)=(10^4,1)$ (solid lines), $(n_\Gamma/n_0,n_0)=(10^{-4},1)$ (dashed lines) or $(n_\Gamma/n_0,n_0)=(10^4,10^2)$ (dot-dashed lines), for model I (red lines) and model II (blue lines). The results for scalar and gravitational wave radiation power are shown by the black and gray solid lines, respectively, the latter is given by $(1-e^2)^{-7/2}(37e^4+292e^2+96)/96$.
 }\label{fig:e}
\end{figure}

The asymptotic limits of $P^\text{(II)}$ in the two models are summarized in Table~\ref{table1}. We also plot the characteristic functions $D_{g'}$ and $D_\Omega$ for varying parameters $(n_{\Gamma}, n_0, e)$ in Fig.~\ref{fig:model_I} and Fig.~\ref{fig:model_II}, with the asymptotic limits indicated by dashed lines. Due to the modification of the scalar propagator, $P^\text{(II)}$ is not simply proportional to $(g')^2$, and in both models it decreases with a sufficiently large $g'$ and inreases with a sufficiently small $g'$. The slopes $\partial_{g'}P^\text{(II)}(g')$ and $\partial_\Omega P^\text{(II)}(\Omega)$ approach constant values for $g'\to 0/\infty$ and $\Omega\to 0/\infty$, which can be read off from Table~\ref{table1}. It can also be seen that the large-$\Omega$ limits of $P^\text{(II)}(\Omega)$ in both models are degenerate with respect to $m$ and $g'$. The radiation is generally enhanced by a larger orbital eccentricity, the enhancement for indirect radiation $P^\text{(II)}$ can be boosted or suppressed relative to the gravitational radiation $P^\text{(0)}$ and scalar radiation $P^\text{(I)}$ depending on the parameters $n_\Gamma/n_0$ and $n_0$, as depicted in Fig.~\ref{fig:e}.

\section{A Specific Scenario and Constraints from Pulsar Binaries}\label{sec3}
The scalar charge may have various physical origins
\cite{Mohanty:1994yi,Hook:2017psm,KumarPoddar:2019jxe,Wong:2019yoc,Seymour:2020yle,Zhang:2023lzt,Lambiase:2024dqe}, in this section we apply our result for the indirect radiation power to a muonophilic scalar with $N$ given by the muon number $N_\mu$, this is one of the minimal models capable of addressing the $(g-2)_\mu$ anomaly \cite{Capdevilla:2021kcf}. Furthermore we allow a nonzero scalar-photon coupling in the form of model I. With the scalar-muon coupling $g\phi \bar \mu \mu$, however, an effective coupling between $\phi$ and the SM photon in the form of model II would arise via a muon loop (see for example \cite{Blinov:2024gcw}, here we neglect the possible UV contribution and take the limit $m\ll m_\mu$), hence the full interaction is
\begin{equation}\label{specific_scenario}
\mathcal{L}_\text{int} = \frac{1}{2}g'\phi A^\mu A_\mu+\frac{1}{4} g''\phi F_{\mu\nu}F^{\mu\nu},
\quad g''\approx\frac{4}{3}\frac{\alpha g}{2\pi m_\mu},
\end{equation}
where $\alpha$ is the fine structure constant and $m_\mu$ is the muon mass. It should be stressed that the second term is a coupling with SM photon while $A_\mu$ in the first term can also be a BSM vector. Being suppressed by a factor of $\Omega^2$, the indirect radiation from the second term is completely negligible compared with the direct scalar radiation for all physical binary parameters.

The muon number density of a neutron star (NS) can be estimated from the beta equilibrium condition and depends on the equation of state of the NS \cite{KumarPoddar:2019ceq,Potekhin:2013qqa}. A conservative estimation is that $N_\mu\sim 10^{55}$ for NS and $N_\mu\sim0$ for the white dwarf (WD) \cite{Garani:2019fpa,Potekhin:2013qqa}. Using the energy flux derived in the last section, we can now place constraints on the couplings (for a given boson mass) from the observational data of pulsar binaries. The conservative dynamics of an inspiralling binary can be described by its effective Lagrangian truncated at certain order in the post-Newtonian (PN) low-velocity expansion\footnote{A PN order of $n$ refers to the correction that scales with $v^{2n}$ relative to the leading term in vacuum GR.}. Since the photon is not coupled to the star, to the leading order the binary's Lagrangian is not affected by the scalar-photon coupling. In the the Newtonian (0PN) regime and for $ma\ll 1$, the scalar potential is unscreened and the orbit is given by $\ddot{\mathbf{X}}=-\frac{\tilde M_\text{tot}}{r^3}\mathbf{X}$, with $\tilde M_\text{tot}=\frac{1}{\mu}\left(M_1M_2+\frac{g^2}{4\pi}N_1N_2\right)$. The orbital energy is
\begin{equation}
E=-\frac{\mu \tilde M_\text{tot}}{2a},
\end{equation}
and $\Omega = \sqrt{\frac{\tilde M_\text{tot}}{a^3}}$. In the adiabatic approximation, the rate of change of orbital period is thus
\begin{equation}
\dot T = -6\pi \left(1+\frac{\frac{g^2}{4\pi}N_1N_2}{M_1M_2}\right)^{-3/2}(M_1M_2)^{-1}(M_1+M_2)^{-1/2}a^{5/2}\left[P+P^{(0)}\right],
\end{equation}
where $P^{(0)}$ is the power of gravitational quadrupole radiation given by Eq.~\eqref{P_0} and $P=P^\text{(I)}+P^\text{(II)}$ is the radiation power due to the ultralight scalar boson given by Eq.~\eqref{P_I} and Eq.~\eqref{model_I}. We obtain the bound as \cite{Seymour:2020yle}
\begin{equation}
\left| (\dot T_\text{b}-\dot T)/\dot T_\text{gw}-\sigma_\text{sys}\right| <2\sigma_\text{stat},
\end{equation}
where $\dot T_\text{gw}=\dot T|_{N_{1,2}=0}$, $\dot T_\text{b}$ is the measured value with fractional standard deviation given by $\sigma_\text{stat}$, and a possible small fractional systematic deviation $\sigma_\text{sys}$ is to be neglected (also we neglect the measurement uncertainty of the binary mass). Here we have neglected the EM radiation due to the possible intrinsic electric charge of the star. The electric charge inside a uniformly magnetized NS can be estimated \cite{1969ApJ...157..869G,1975ApJ...196...51R} as $q_\text{NS}\approx (2/3)\omega B_\text{P}{R^3}/c$, where $R$, $\omega$, $B_\text{P}$ are the radius, spin angular velocity, and the surface dipole magnetic field (in Gaussian units) of the NS. Taking the canonical parameters $R=10\,\text{km}$, $\omega=10^3\,\text{Hz}$ and a strong magnetic field $B_\text{P}=10^{14}\,\text{G}$ gives $q_\text{NS}\sim 10^{14}\,\text{C}$ (in SI units), this amounts to be the coupling strength $g=10^{-23}$ with massless vector for $q=g^{-1}\sqrt{\mu_0 c/\hbar}\,q_\text{NS}\sim 10^{55}$ (see \ref{appendix_2}). So it is reasonable to neglect the electric charge of the NS (see also \cite{PhysRevD.49.6892}; the possibilities of probing an ectrophilic scalar interacting with the electrons in NS was explored in \cite{Lambiase:2024dqe}). Due to the scalar-photon or or pseudoscalar-photon coupling, the oscillating EM fields of a rotating NS can generate scalar radiation \cite{Astashenkov:2023wmt,Khelashvili:2024sup} (and also GWs \cite{Contopoulos:2023iom}, from the minimal graviton-photon coupling), but like any intrinsic EM process of the star (such as the magnetic dipole radiation of a rotating NS), it does not backreact on the binary's orbital motion. The binary can be affected by its environment, e.g., the gravitational dynamical friction in a dark matter background, but its effect appears to be negligible if the dark matter density $\rho_\text{DM}\ll 10^5\,\text{GeV}/\text{cm}^3$ \cite{Pani:2015qhr} (this estimation was made for the CDM, see \cite{Wong:2019yoc,Blas:2019hxz,Brax:2024yqh,Koo:2023gfm,Bromley:2023yfi,PhysRevLett.132.211401} for discussions on the effects of ultralight scalar dark matter). Here we do not assume $\phi$ to constitute all the dark matter and neglect its background value.

We examine two NS-WD binaries, for which the scalar contribution to the binding energy can be neglected. The observational data of their orbital parameters are listed in Table~\ref{table2}. Since the NS radius $R_\text{NS}\sim 10\,\text{km}\approx 5\times 10^{10}\,\text{eV}^{-1}$ is much smaller than the orbital period and the reduced Compton wavelength $m^{-1}$ in the considered mass range, the point charge approximation is valid. The binary separation is also sufficiently small so that the dipole approximation is valid; these are basically the same conditions under which the quadrupole formula can be used to describe the GW radiation.

\begin{table}[h!]
\renewcommand{\arraystretch}{1.5}
    \begin{center}
        \begin{tabular}{ccc}
        \hline
         \textbf{Parameters} & \textbf{PSR J1141-6545} \cite{Bhat:2008ck} & \textbf{PSR J1738+0333} \cite{Freire:2012mg}
         \\
         \hline
         $M_1$ $(M_{\odot})$ &1.27(1)& 1.46(6) \\
         $M_2$ $(M_{\odot})$ &1.02(1)& 0.181(8) \\ 
         $e$ &0.171884(2)& $3.4(11) \times 10^{-7}$ \\
         $\dot{T}_\text{b}$ &$-0.403(25)\times10^{-12}$& $ -2.59(32) \times 10^{-14}$ \\
         $\Omega=2\pi/T_\text{b}$ (eV) &$2.421 \times 10^{-19}$&$1.349 \times 10^{-19}$\\
         \hline
        \end{tabular}
    \end{center}
    \caption{The orbital parameters of PSR J1141-6545 and PSR J1738+0333, figures in parenthesis are the 1$\sigma$ uncertainties in the last quoted digit. Note that $\dot{T}_\text{b}=\dot{T}_\text{b}^\text{obs}-\dot{T}_\text{b}^\text{acc}-\dot{T}_\text{b}^\text{shk}$, where $\dot{T}_\text{b}^\text{obs}$ is the apparent decay rate, and $\dot{T}_\text{b}^\text{acc}$, $\dot{T}_\text{b}^\text{shk}$ the corrections due to kinematic effects.}
    \label{table2}
\end{table}

\begin{figure}[hbt]
	\centering
	\includegraphics[width=0.46\textwidth]{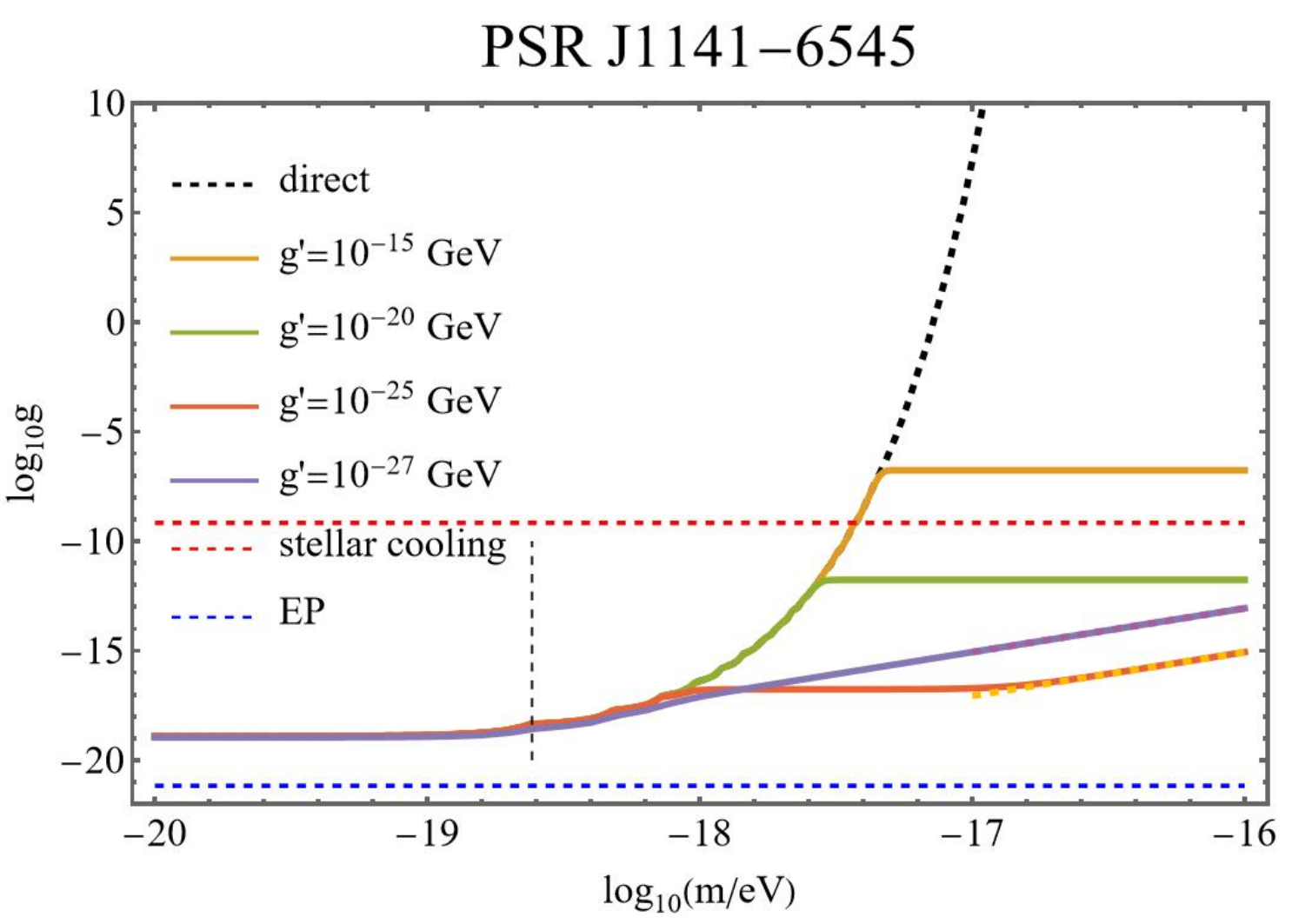}
 \qquad
    \includegraphics[width=0.46\textwidth]{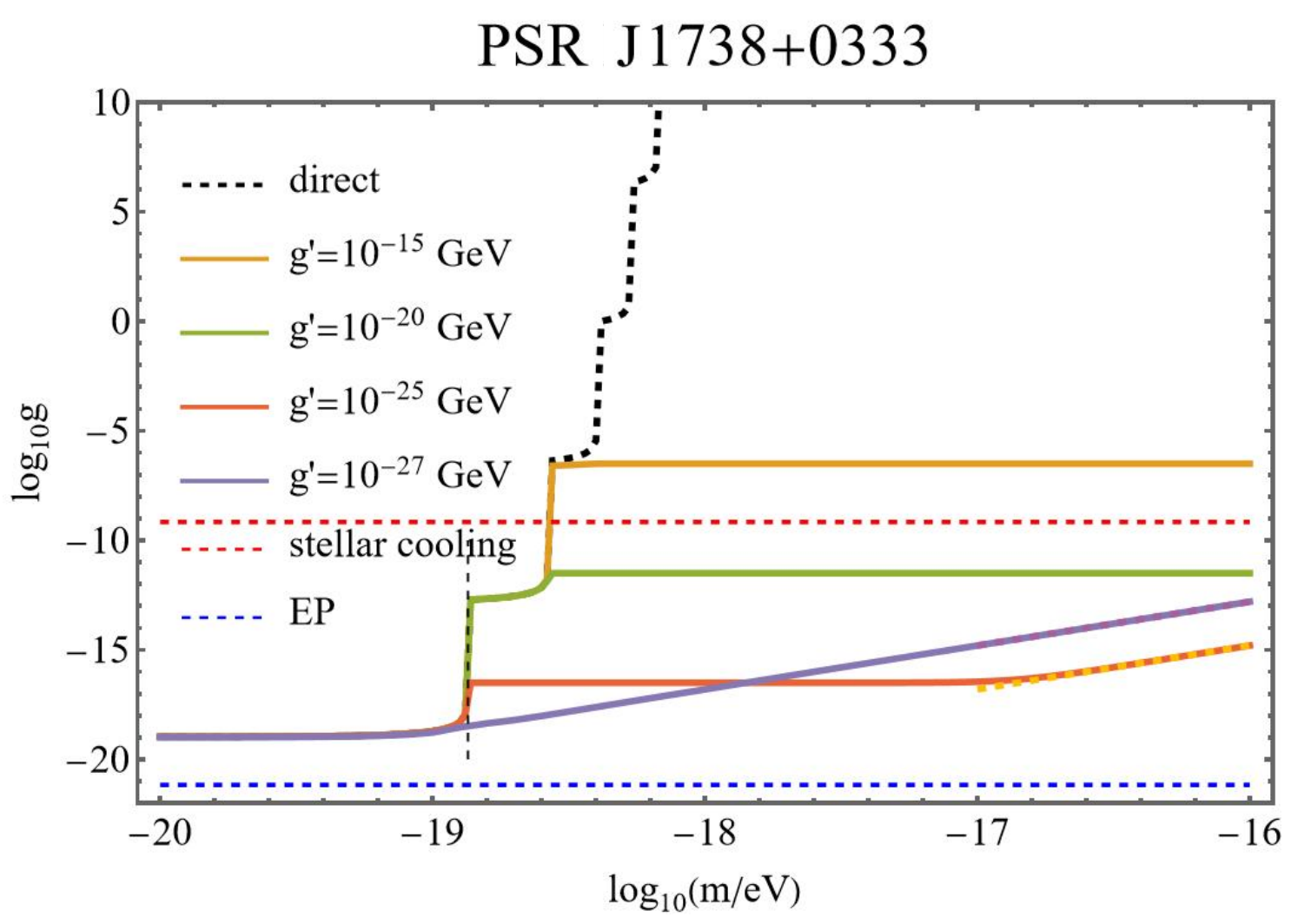}
	\caption{Constraints on $g$ and $m$ from two PSR-WD binaries for given values of $g'$. The solid (dot-dashed) line corresponds to the result with (without) indirect radiation, the dashed line corresponds to the $g'\to 0$ limit. The critical mass $m=\Omega$ is indicated by the vertical line.}\label{fig_3}
\end{figure}

The constraints on $g$ for various values of $g'$ are depicted in Fig.~\ref{fig_3}. As can be seen, the indirect radiation starts to dominate for $m\gtrsim \Omega$. For $g'\lesssim 10^{-25}\,\text{GeV}$, the constraint on $g$ weakens as $g'$ decreases, since the indirect radiation power peaks at $g' \approx 10^{-25}\,\text{GeV}$. If $g'$ is small enough, the small-$g'$ limit presented in Table~\ref{table1} appears to be good approximation for a large enough boson mass. The constraint on $g$ is approximately constant when the indirect radiation dominates if $g'$ is sufficiently large, since the large-$g'$ limit of $P^\text{(II)}$ is independent of the boson mass (see Table~\ref{table1}).

The induced $\phi F^2$ coupling from the scalar-muon coupling (second term in Eq.~\eqref{specific_scenario}), although being negligible for the radiative dynamics of the binary, can be probed by a variety of other experiments and observations \cite{Antypas:2022asj}, which then leads to some stringent constraints on $g$ when $\phi$ is ultralight. For example, we show the constraints from equivalence principle (EP) tests \cite{Hees:2018fpg} and stellar cooling \cite{raffelt1996stars} in Fig.~\ref{fig_3}. It turns out that in the present scenario, the current constraints on $g$ from PSR-WD binaries is considerably weaker than that derived from the EP tests, even in the massless limit where the additional radiation power is dominated by $P^\text{(I)}$ and takes its maximum value. The constraints can nonetheless be improved in the future when the orbital decay of PSR binaries is measured with a higher precision.

\section{Summary and Discussion}\label{sec4}
We have investigated the scenario in which a massive scalar boson couples simultaneously to a massless photon and the matter constituents of compact stars, so that a binary could generate both the scalar radiation and the indirect EM radiation mediated by the scalar. In the case of the vector being SM photon, such an indirect EM radiation would be nearly unobservable due to its extremely low frequency ($\omega\sim \Omega_n$), but it may lead to detectable signals from the induced secondary EM processes in the interstellar medium \cite{PhysRevD.108.083018}, such as the synchrotron radiation. The focus of this paper is the dipole energy flux from a charged binary in elliptical orbit, but the calculation can be extended straightforwardly to the quadruple order and hyperbolic orbit, as demonstrated in the appendices. It is also possible, though less straightforward, to obtain the angular momentum radiation flux of a Keplerian binary\footnote{See \ref{appendix_3} for the angular momentum flux of direct radiation.}, from which the evolution of orbital eccentricity can be derived. But the conservative dynamics will be more complicated if the scalar-mediated force is non-negligible but is partially screened by the scalar mass\footnote{Note that a coupling between $\phi^n$ and the body's worldline with $n\ge 2$ alone will not modify the binary's conservative or radiative dynamics in the classical level, the computation of scalar radiation power in this case is similar to that of indirect process in Sec.~\ref{sec2}.}. In this case, an orbital parametrization taking into account the Yukawa potential is needed to derive the radiation fluxes for non-circular orbits. We leave these issues for future studies.

\section*{Acknowledgments}
We thank Dr. Yong Tang for helpful discussions and suggestions. We also thank the anonymous referee for useful comments.

\appendix
\section{Scalar Quadrupole Radiation}\label{appendix_1}
In this appendix, we consider the quadrupole radiation from an elliptical binary with scalar charges, assuming that the scalar charges are conserved. In the momentum space, the scalar charge density can be expanded as
\begin{equation}
\begin{aligned}
n(\Omega_n,\mathbf{k})
&=
\frac1T\int_0^Tdt\int d^3x\, e^{-i\mathbf{k}\cdot\mathbf{x}+i\Omega_n t}n(t,\mathbf{x})
\\
&=
\frac1T\int_0^Tdt\sum_{I=1,2}\int d^3x\, e^{-i\mathbf{k}\cdot\mathbf{x}+i\Omega_n t}N_I\delta^3(\mathbf{x}-\mathbf{X}_I(t))
\\
&=
\sum_{\ell=1}^\infty\frac1T\int_0^Tdt\sum_{I=1,2}N_I\,\left[\frac{(-i)^{\ell}}{\ell!}\prod_{l=1}^{\ell}k_{i_l}X_{Ii_l}\right]\,e^{i\Omega_nt}
\\
&\equiv \sum_\ell n^{(\ell)}(\Omega_n,\mathbf{k})
,
\end{aligned}
\end{equation}
where the $\ell=1$ term alone is the source of dipole radiation (see Eq.~\eqref{dipole}), it dominates over the $\ell=2$ term if the charge-to-mass ratio difference $D$ is sufficiently large, in which case the calculation of quadrupole radiation is unnecessary, since it is suppressed by a factor of $v^2\sim a^2\Omega^2$ relative to the dipole radiation. Therefore, in the following we focus on a vanishing dipole moment in the binary's center of mass frame (so that $N_1/M_1=N_2/M_2$), the radiation is then dominated by the $\ell=2$ term:
\begin{align}
n(\Omega_n,\mathbf{k})&\approx n^{(2)}(\Omega_n,\mathbf{k})=
-k_ik_jI_{ij},
\\
I_{ij}&\equiv\frac{\tilde N}{2T}\int_0^Tdt\,X_i(t)X_j(t)\,e^{i\Omega_nt},
\\
\tilde N&\equiv\frac{N_1M_1^2+N_2M_2^2}{(M_1+M_2)^2}.
\end{align}
For $N_1/M_1=N_2/M_2=\kappa$, $\tilde N=\mu \kappa$. Together with Eq.~\eqref{radiation_power_formula} the power of quadrupole radiation can be derived for a given process.

For simplicity, here we consider only the direct scalar radiation given by the amplitude $i\mathcal{M}_n=ign(\Omega_n,\mathbf{k})$, straightforward calculation leads to
\begin{equation}
P_\text{quad}^\text{(scalar)}=\sum_{n\ge n_0}^\infty g^2a^4\tilde N^2\Omega ^6\left(1-\frac{n_0^2}{n^2}\right)^{5/2}f_\text{quad}^\text{(scalar)}(n,e),
\end{equation}
with $n_0=m/\Omega$, and
\begin{equation}
\begin{aligned}
f_\text{quad}^\text{(scalar)}(n,e)=&\,n^2\big[
e^6 n^2 \left(J_{n-1}^2-2 J_{n+1} J_{n-1}-4 J_n^2+J_{n+1}^2\right)
\\
&+e^5 n (6 J_n J_{n+1}-6 J_{n-1} J_n)+e n (8 J_n J_{n+1}-8 J_{n-1} J_n)
\\
&+e^4 n^2 \left(-2 J_{n-1}^2+4 J_{n+1} J_{n-1}+12 J_n^2-2 J_{n+1}^2\right)
\\
&+e^4\left(-J_{n-1}^2+3 J_n^2-J_{n+1}^2+2 J_{n-1} J_{n+1}\right)
\\
&+e^3 n (14 J_{n-1} J_n-14 J_n J_{n+1})+4 n^2 J_n^2+4 J_n^2
\\
&+e^2 n^2 \left(J_{n-1}^2-2 J_{n+1} J_{n-1}-12 J_n^2+J_{n+1}^2\right)
\\
&+e^2\left(J_{n-1}^2-4 J_n^2+J_{n+1}^2-2 J_{n-1} J_{n+1}\right)\big]/(30\pi e^4)
,
\end{aligned}
\end{equation}
where $J_m\equiv J_m(ne)$. For $e=0$, only the $n=2$ term contributes to the radiation power: $P_\text{quad}^\text{(scalar)}(e=0)=\frac{4}{15\pi}g^2 a^4 \tilde N^2\Omega ^6\left(1-\frac{n_0^2}{4}\right)^{5/2}$, which is same with the result derived using an EFT approach in \cite{Huang:2018pbu} for circular orbit. In the massless limit, the infinite series can be evaluated analytically and we obtain
\begin{equation}
P_\text{quad}^\text{(scalar)}(m=0)=
\frac{4}{15\pi}g^2 a^4 \tilde N^2\Omega ^6
\frac{51 e^4+396 e^2+128}{128 \left(1-e^2\right)^{7/2}},
\end{equation}
as can be easily checked, this is compatible with the energy density of the quadrupole radiation field given by the Klein-Gordon equation $(-\partial_t^2+\nabla^2)\phi=-gn$ (see for example \cite{10.1119/1.13627}).

\section{Vector Dipole and Quadrupole Radiation}\label{appendix_2}
The dipole and quadrupole radiation power of massive vector field from an elliptical binary with vector charge $q_{1,2}$ can be analogously derived, the relevant Lagrangian is given by
\begin{equation}
\mathcal{L}\supset \frac{1}{2}m^2\mathcal{A}_\mu \mathcal{A}^\mu-\frac{1}{4}\mathcal{F}^{\mu\nu}\mathcal{F}_{\mu\nu}+g J_\mu \mathcal{A}^\mu,
\end{equation}
with $\mathcal{F}_{\mu\nu}=\partial_\mu \mathcal{A}_\nu -\partial_\nu \mathcal{A}_\mu$, where $\mathcal{A}_\mu$ is a vector field with mass $m$, and $J^\mu=\sum_{I=1,2} q_I(1,\dot{\mathbf{X}}_I)\,\delta^3(\mathbf{x}-\mathbf{X}_I(t))$ is the source charge current density. The amplitude of vector radiation is $i\mathcal{M}_n=igJ^\mu(\Omega_n,\mathbf{k})\bar\epsilon_\mu^{(\lambda)}(\mathbf{k})$, where $\epsilon_\mu^{(\lambda)}(\mathbf{k})$ is the normalized polarization vector with $\epsilon^{(\lambda)}\cdot \bar\epsilon^{(\lambda')}=\delta_{\lambda,\lambda'}$. With the help of charge conservation $\partial_\mu J^\mu=0$, and upon performing the polarization sum, we obtain
\begin{equation}
\sum_{\lambda_1,\lambda_2}|\mathcal{M}_n|^2=
g^2\left(-\frac{k_{i}k_{j}}{\Omega_n^{2}}J_{i}\bar{J}_{j}+J_{i}\bar{J}_{i}
\right).
\end{equation}
The current density can be expanded as
\begin{equation}
\begin{aligned}
J_i(\Omega_n,\mathbf{k})
&=
\frac1T\int_0^Tdt\int d^3x\, e^{-i\mathbf{k}\cdot\mathbf{x}+i\Omega_n t}J_i(t,\mathbf{x})
\\
&=
\frac1T\int_0^Tdt\sum_{I=1,2}\int d^3x\, e^{-i\mathbf{k}\cdot\mathbf{x}+i\Omega_n t}q_I\,\dot X_{Ii}\,\delta^3(\mathbf{x}-\mathbf{X}_I(t))
\\
&=
\frac1T\int_0^Tdt\sum_{I=1,2}q_I\,\left[
\dot X_{Ii}
+
\sum_{\ell=1}^\infty\frac{(-i)^{\ell}}{\ell!}\prod_{l=1}^{\ell}k_{i_l}X_{Ii_l}\dot X_{Ii}\right]\,e^{i\Omega_nt}
\\
&\equiv \sum_{\ell=0}^\infty J_i^{(\ell)}(\Omega_n,\mathbf{k})
,
\end{aligned}
\end{equation}
where the $\ell=0$ term alone is the source of electric dipole radiation, and $\ell=1$ term alone gives rise to the electric quadrupole and magnetic dipole radiation \cite{PhysRevD.49.6892}. But for a charged Keplerian binary, the magnetic dipole radiation vanishes, since the magnetic moment is proportional to the nearly conserved angular momentum of the binary. For the electric dipole radiation, straightforward calculation gives that (see also \cite{PhysRevD.49.6892,KumarPoddar:2019ceq})
\begin{equation}
P_\text{dip}^\text{(vector)}=\sum_{n\ge n_0}^\infty\frac{1}{6\pi}g^2a^2{\mu}^2D^2\Omega^4n^2\left[(J_n')^2+\frac{1-e^2}{e^2}(J_n)^2\right]\left(1-\frac{n_0^2}{n^2}\right)^{1/2}\left(2+\frac{n_0^2}{n^2}\right),
\end{equation}
with $D=\frac{q_1}{M_1}-\frac{q_2}{M_2}$. In the massless limit,
\begin{equation}
P_\text{dip}^\text{(vector)}=
\frac{1}{6\pi}g^2a^2\mu^2D^2\Omega^4\frac{(1+e^2/2)}{(1-e^2)^{5/2}}=2P^\text{(I)}.
\end{equation}
For the electric quadrupole radiation, straightforward calculation leads to
\begin{equation}
P_\text{quad}^\text{(vector)}=\sum_{n\ge n_0}^\infty g^2a^4\tilde q^2\Omega^6\left(1-\frac{n_0^2}{n^2}\right)^{3/2}f_\text{quad}^\text{(vector)}(n,e),
\end{equation}
with $\tilde q=\frac{q_1M_1^2+q_2M_2^2}{(M_1+M_2)^2}$, and
\begin{equation}
\begin{aligned}
f_\text{quad}^\text{(vector)}(n,e)=\,&\Big\{
2(3n^4+2n_0^2n^2) ( J_{n-1}^2-4 J_n^2+ J_{n+1}^2-2 J_{n-1} J_{n+1})e^6
\\
&+12(3n^3 + 2n_0^2n)(J_n J_{n+1} - J_{n-1} J_n ) e^5
\\
&+\Big[
4(3n^4+2n_0^2n^2)
(- J_{n-1}^2 +6 J_n^2 -J_{n+1}^2 +2 J_{n-1} J_{n+1})
\\
&+n^2(8 J_n^2 - 6J_{n-1}^2 - 6J_{n+1}^2 +12 J_{n-1} J_{n+1})
\\
&+n_0^2(12J_n^2-4J_{n-1}^2-4 J_{n+1}^2+8J_{n-1} J_{n+1})\Big] e^4
\\
&-28(3 n^3+2n_0^2n)(J_n J_{n+1}-J_{n-1} J_n) e^3
\\
&+\Big[2(3n^4+2n_0^2n^2)( J_{n-1}^2 -12 J_n^2 + J_{n+1}^2 -2 J_{n-1} J_{n+1})
\\
&+2(3n^2+2n_0^2)( J_{n-1}^2 -4 J_n^2 + J_{n+1}^2-2J_{n-1} J_{n+1}) \Big] e^2
\\
&+16(3 n^3+2n_0^2n)(J_n J_{n+1} - J_{n-1} J_n) e
\\
&+8(3n^2+2n_0^2) (n^2+1) J_n^2
\Big\}/(120\pi e^4)
.
\end{aligned}
\end{equation}

In the massless limit, the infinite series can also be evaluated analytically and we obtain
\begin{equation}
P_\text{quad}^\text{(vector)}(m=0)=
\frac{2}{5\pi}g^2 a^4 \tilde q^2\Omega ^6
\frac{37 e^4+292 e^2+96}{96 \left(1-e^2\right)^{7/2}}
=
\frac{g^2\tilde q^2}{16\pi \mu^2}
P^{(0)}
,
\end{equation}
this is compatible with the energy density of the electric quadrupole radiation field given by the Maxwell equation $\partial_\mu F^{\mu\nu}=-gJ^\nu$ (see for example \cite{Christiansen:2020pnv}). For $e=0$, the radiation power is
\begin{align}
P_\text{dip}^\text{(vector)}(e=0)&=
\frac{1}{6\pi}g^2 a^2\mu^2D^2\Omega ^4\left(1-n_0^2\right)^{1/2}\left(1+\frac{n_0^2}{2}\right)
,
\\
P_\text{quad}^\text{(vector)}(e=0)&=
\frac{2}{5\pi}g^2 a^4 \tilde q^2\Omega ^6\left(1-\frac{n_0^2}{4}\right)^{3/2}\left(1+\frac{n_0^2}{6}\right)
.
\end{align}

Incidentally, we can consider a massive dark photon field $\mathcal{A}_\mu$ sourced by $J_\mu$ with kinetic mixing $\alpha$ and mass mixing $\chi$ to the SM photon $A_\mu$, such a system is described by the Lagrangian:
\begin{equation}
\mathcal{L}\supset \frac{1}{2}m^2\mathcal{A}_\mu \mathcal{A}^\mu-\frac{1}{4}\mathcal{F}^{\mu\nu}\mathcal{F}_{\mu\nu}+g J_\mu \mathcal{A}^\mu
+\frac{\sin\alpha}{2} F_{\mu\nu}\mathcal{F}^{\mu\nu}+\chi m^2A_\mu \mathcal{A}^\mu-\frac{1}{4}F^{\mu\nu}F_{\mu\nu}.
\end{equation}
For $\chi=0$, through a change of basis: $\mathcal{A}_\mu\to\frac{1}{\cos\alpha}\mathcal{A}_\mu$, $A_\mu\to A_\mu+\tan\alpha\mathcal{A}_\mu$, the result is a decoupled pair of $A_\mu$ and $\mathcal{A}_\mu$ fields with the latter being sourced by an enlarged current $J_\mu'=J_\mu/\cos\alpha$ and with an enlarged mass $m'=m/\cos\alpha$. For $\alpha=0$, the indirect radiation of $A_\mu$  due to the mass mixing turns out to be equivalent to the radiation from a source current $g\chi J_\mu$ (so this is a classical process).

\section{Angular Momentum Flux}\label{appendix_3}
In this appendix we compute the angular momentum flux associated with the dipole radiation of massive scalar and vector fields from a charged binary in elliptical orbit, the results are Eq.~\eqref{J_flux_scalar} and \eqref{J_flux_vector}, respectively. To this end, the radiation field has to be obtained explicitly, which means that we have to resort to the traditional approach (see for example \cite{GW_volume_1}).

Consider first the case of scalar charge, the radiation field is \cite{PhysRevD.49.6892}
\begin{equation}
\phi(t,\mathbf{x}=r\mathbf{n})=\frac{g}{4\pi r}\sum_{|n|\ge n_0}\left[n(\Omega_n,\mathbf{k})\,e^{i(\mathbf{k}\cdot\mathbf{x}-\Omega_nt)}\right]_{\mathbf{k}=\mathbf{k}^{(n)}\equiv k_n\mathbf{n}}
,
\end{equation}
with $|\mathbf{n}|=1$ and $k_n\equiv \Omega_n\sqrt{1-(m/\Omega_n)^2}$. Same as the massless case, the volume density of angular momentum is $j_i=-\epsilon_{ikl}\dot{\phi}x^{k}\partial_{l}\phi$ (which is purely orbital), under the time average:
\begin{align}
\frac{-\langle j_i \rangle}{\left(\frac{g}{4\pi r}\right)^2}&=\frac{1}{T}\int_0^T dt\,\epsilon_{ikl}x_k\sum_{n,m}(-i\Omega_n) n(\Omega_n,\mathbf{k}^{(n)})[\partial_ln(\Omega_m,\mathbf{k}^{(m)})]\,
e^{i(k_n+k_m)\mathbf{n}\cdot\mathbf{x}}e^{-i(n+m)\Omega t}\nonumber
\\
&=\epsilon_{ikl}x_k\sum_{n}\Omega_n n(\Omega_n,\mathbf{k}^{(n)})[-i\partial_l\bar n(\Omega_n,\mathbf{k}^{(n)})]
.
\end{align}
The time-averaged angular momentum flux is then given by
\begin{equation}
\tau_i=-r^2\int d\Omega_\mathbf{n}\left(\frac{g}{4\pi r}\right)^2\epsilon_{ikl}x_k\sum_{n}\Omega_n n(\Omega_n,\mathbf{k}^{(n)})[-i\partial_l\bar n(\Omega_n,\mathbf{k}^{(n)})]\,v_g^{(n)}
,
\end{equation}
where $v_g^{(n)}=\frac{k_n}{\Omega_n}=\sqrt{1-(m/\Omega_n)^2}$ is the group velocity of the (outgoing) $\mathbf{k}^{(n)}$-mode.

For the dipole radiation, we take $n(\Omega_n,\mathbf{k})=i(a\mu D)(-i\mathbf{k})\cdot\mathbf{j}_n=(a\mu D)k_n\mathbf{n}\cdot\mathbf{j}_n$, with $\mathbf{j}_n$ given by Eq.~\eqref{j_n}. Using $x_k\partial_l n_j=n_k(\delta_{lj}-n_ln_j)$, we obtain\footnote{Incidentally, $P/\Omega_n-\tau\propto [(J_n')^2+\frac{1-e^2}{e^2}(J_n)^2]-\frac{2(1-e^2)^{1/2}}{e}J_n'J_n=[J_n'-\frac{(1-e^2)^{1/2}}{e}J_n]^2$.}
\begin{align}
\boldsymbol{\tau}=-\dot{\mathbf{J}}
&=\frac{g^2}{6\pi}(a\mu D)^2\sum_{n\ge n_0} (-i)\,k_n^3\,\mathbf{j}_n\times \bar{\mathbf{j}}_n
\\
&=\frac{g^2}{3\pi}(a\mu D)^2\Omega^3\sum_{n\ge n_0}\frac{(1-e^2)^{1/2}}{e}nJ_n'J_n\,\left(1-\frac{n_0^2}{n^2}\right)^{3/2}\hat{\mathbf{J}}\equiv \tau \hat{\mathbf{J}}\label{J_flux_scalar}
,
\end{align}
where $\hat{\mathbf{J}}$ is a unit vector parallel to the orbital angular momentum $\mathbf{J}=J\hat{\mathbf{J}}$ of the binary. As a consistency check, for circular orbit: $\tau(e=0)=\tau_{n=1}=\frac{g^2}{12\pi}(a\mu D)^2\Omega^3(1-n_0^2)^{3/2}=P/\Omega$ (the energy flux is given by Eq.~\eqref{P_I}); in the massless limit:
\begin{equation}
\tau=\frac{g^2}{3\pi}(a\mu D)^2\Omega^3\sum_{n=1}^\infty\frac{(1-e^2)^{1/2}}{e}nJ_n'J_n
=\frac{g^2}{12\pi}(a\mu D)^2\Omega^3(1-e^2)^{-1}
.
\end{equation}
which matches the result in \cite{Cardoso:2020iji}.

In the case of vector charge, we obtain analogously for the electric dipole radiation:
\begin{align}
\boldsymbol{\tau}
=\frac{g^2}{3\pi}(a\mu D)^2\Omega^3\sum_{n\ge n_0}\frac{(1-e^2)^{1/2}}{e}nJ_n'J_n\,\left(1-\frac{n_0^2}{n^2}\right)^{1/2}\left(2+\frac{n_0^2}{n^2}\right)\hat{\mathbf{J}}\label{J_flux_vector}
.
\end{align}
Eq.~\eqref{J_flux_vector} can be understood as follows: the angular momentum carried by the two transverse modes $\mathcal{A}_i^\text{T}=(\delta_{ij}-n_in_j)\mathcal{A}_j$ is largely same as the massless case, only with an extra factor $\left(1-\frac{n_0^2}{n^2}\right)^{1/2}$ from the modified group velocity, hence $\boldsymbol{\tau}_n^\text{T}=\left(1-\frac{n_0^2}{n^2}\right)^{1/2}\boldsymbol{\tau}_n(m=0)$. The longitudinal mode is obtained from the radial projection $\mathcal{A}^\text{L}_i=n_in_j\mathcal{A}_j$, and is similar to the scalar dipole radiation, including the normalization factor\footnote{In the Fourier decomposition:\[\boldsymbol{\mathcal{A}}(t,\mathbf{x})=\sum_{\lambda=\pm, \parallel}\int_\mathbf{k}f_{\lambda,\mathbf{k}}(t)\,\boldsymbol{\epsilon}^{(\lambda)}(\mathbf{k})\,e^{i\mathbf{k}\cdot\mathbf{x}},
\quad
|\boldsymbol{\epsilon}^{(\lambda)}|=1,
\quad
\boldsymbol{\epsilon}^{(\parallel)}=\mathbf{k}/|\mathbf{k}|,
\quad
\boldsymbol{\epsilon}^{(\parallel)}\cdot \boldsymbol{\epsilon}^{(\pm)}=0
,
\]
(where $\lambda=\pm,\parallel$ correspond to the transverse and longitudinal $\mathbf{k}$-modes, respectively) the free Proca Lagrangian in flat spacetime reads:$
\int d^3x\left(\frac{1}{2}m^2\mathcal{A}_\mu\mathcal{A}^\mu-\frac{1}{4}\mathcal{F}^{\mu\nu}\mathcal{F}_{\mu\nu}\right)
=\int_\mathbf{k}\left\{\sum_{\lambda=\pm}\frac{1}{2}\left[|\dot{f}_{\lambda,\mathbf{k}}|^2-\omega_k^2|f_{\lambda,\mathbf{k}}|^2\right]+\frac{m^2/\omega_k^2}{2}\left[|\dot{f}_{\|,\mathbf{k}}|^2-\omega_k^2|f_{\|,\mathbf{k}}|^2\right]\right\}.
$
} $m^2/\Omega_n^2=n_0^2/n^2$, its contribution to the flux is therefore $\boldsymbol{\tau}^\text{L}=\boldsymbol{\tau}^\text{T}\frac{n_0^2}{2n^2}$. In similar ways one can derive the angular momentum flux associated with the quadrupole radiation of massive scalar and vector fields.

The angular momentum flux of the indirect radiation cannot be computed in this approach, since the associated radiation field is non-classical (i.e., not in the coherent state). However, based on the results for the energy and angular momentum flux of the direct radiation, as well as the consistency between energy and angular momentum flux in the case of circular orbit, a plausible guess is that the dipolar angular momentum flux of the indirection radiation can also be obtained from its enegy flux via the replacement: $(J_n')^2+\frac{1-e^2}{e^2}(J_n)^2\to \frac{2(1-e^2)^{1/2}}{en}J_n'J_n$, for each harmonic number $n$.

\section{Radiation from Hyperbolic Orbit}\label{appendix_4}
Besides the radiation from bound orbits, there are also possibilities of bremsstrahlung radiation from an unbound orbit, which at the Newtonian order can be parameterized by the eccentric anomaly $\xi\in(-\infty,\infty)$ as
\begin{equation}
X(t)=a(e-\cosh\xi),\quad Y(t)=b\sinh\xi,\quad\Omega t=e\sinh\xi-\xi.
\end{equation}
with $Z(t)=0$, eccentricity $e>1$ and $\Omega=\sqrt{M_\text{tot}/a^3}$, if we neglect the modifications to the binary's binding energy. The calculation of radiation power is same as the elliptical orbit despite that in the present case $n\in \mathbb{R}_{\ge 0}$ (also the Fourier integration $\frac{1}{T}\int_0^T dt$ is replaced by $\int_{-\infty}^{\infty} dt$), then $P_n$ represents the spectral density of the total radiated energy at the frequency $\omega=n\Omega$, viz. $\Delta E=\int_{-\infty}^{\infty} dt\,P=\frac{1}{2\pi}\int_0^\infty d\omega\,P_n$.

For gravitational quadrupole radiation, we obtain (see also \cite{DeVittori:2012da,Garcia-Bellido:2017knh,Hait:2022ukn})
\begin{equation}
\begin{aligned}
    P_n=&\,\frac{32\pi^2}{20} a^4 \mu^2 n^4\Omega^4 \bigg\{\left[\frac{2 \left(e^2-1\right)}{e^2 n^2}+\frac{2 \left(e^2-1\right)^2}{e^2}\right] |H'_{i n}|^2+\\
&\,\left[\frac{2 \left(e^4-3 e^2+3\right)}{3 e^4 n^2}+\frac{2 \left(3 e^6-9 e^4+9 e^2-3\right)}{3 e^4}\right] |H_{i n}|^2\bigg\},
\end{aligned}
\end{equation}
with $H_{in}\equiv H_{in}^{(1)}(ine)$, $H_{in}'\equiv \frac{dH_{in}^{(1)}(z)}{dz}|_{z=ine}$, where $H^{(1)}_n(z)$ is the Hankel function of the first kind. In particular, since $\lim_{n\to 0}H_{in}=\frac{2i}{\pi}\ln (ne)$ and $\lim_{n\to 0}H'_{in}=\frac{2}{\pi ne}$, the zero mode radiation density is $P_0=\frac{32}{5} a^4 \mu^2 \Omega^4\frac{2 \left(e^2-1\right)}{e^4}$. The total radiated energy (see also \cite{DeVittori:2012da}) and angular momentum are
\begin{align}
\Delta E&=\frac{2}{45}a^4 \mu ^2 \Omega ^5\frac{\left(673 e^2+602\right) \sqrt{e^2-1} + 3 \left(37 e^4+292 e^2+96\right) \arccos\left(-\frac{1}{e}\right)}{\left(e^2-1\right)^{7/2}},
\\
\Delta J&=\frac{8}{5}a^4 \mu ^2 \Omega ^4\frac{\left(2 e^2+13\right) \sqrt{e^2-1} + \left(7 e^2+8\right) \arccos\left(-\frac{1}{e}\right)}{\left(e^2-1\right)^2}.
\end{align}
The parabolic limit is obtained by the replacement $e\to 1$ and $e-1\to p/(2a)$, where $p$ is the semi-latus rectum of the parabolic orbit.

For a binary with scalar charges, the spectrum of dipolar scalar radiation is given by
\begin{align}
P_n&=\frac{\pi}{6}g^2 a^2\mu^2D^2\Omega^2n^2
f_\text{dip}(n,e)
\left(1-\frac{n_0^2}{n^2}\right)^{3/2}
,
\\
P_0(m=0)&=\frac{2}{3\pi}g^2 a^2\mu^2D^2\Omega^2e^{-2},
\\
\Delta E(m=0)&=
\frac{a^2 Q^2 \Omega ^3 }{12 \pi}\frac{3\sqrt{e^2-1}+\left(e^2+2\right) \arccos\left(-\frac{1}{e}\right)}{\left(e^2-1\right)^{5/2}},
\\
\Delta J(m=0)&=
\frac{a^2 Q^2 \Omega ^2 }{12 \pi}
\frac{\sqrt{e^2-1}+\arccos\left(-\frac{1}{e}\right)}{e^2-1}
.
\end{align}
where $f_\text{dip}(n,e)\equiv\left(1-\frac{1}{e^2}\right) |H_{in}|^2+|H'_{i n}|^2$. For a binary with vector charges, the  spectrum of dipolar vector radiation is
\begin{align}
P_n&=\frac{\pi}{6}g^2 a^2\mu^2D^2\Omega^2n^2
f_\text{dip}(n,e)
\left(1-\frac{n_0^2}{n^2}\right)^{1/2}\left(2+\frac{n_0^2}{n^2}\right)
.
\end{align}
The non-vanishing of $P_0$ is a signature of the memory effect (the difference between the field values at the asymptotic past and future, as viewed by a distant observer), which appears only in the massless case ($m=0$). The time-domain waveforms can be easily computed in the massless case, one can also derive the frequency-domain waveforms in the massive case.\footnote{This is the Newtonian-order waveform. Apart from the PN corrections, the post-Minkowskian waveform can be computed using the approach of worldline quantum field theory \cite{Bhattacharyya:2024aeq}.}

Finally, we give the spectrum of the indirect scalar-mediated EM radiation considered in the main text (at the dipole order), which is
\begin{equation}
P_n=\frac{1}{96{\pi}^3}g^2(g')^2a^2{\mu}^2D^2n^{-1}
f_\text{dip}(n,e)
\int_{0}^{n}dx\,F(x),
\end{equation}
for model I (with $F(x)$ given by Eq.~\eqref{F_model_I}) and
\begin{equation}
P_n=\frac{\Omega^4}{48{\pi}^3}g^2(g')^2a^2{\mu}^2D^2n^{-1}
f_\text{dip}(n,e)
\int_{0}^{n}dx\,F(x),
\end{equation}
for model II (with $F(x)$ given by Eq.~\eqref{F_model_II}).

\bibliographystyle{elsarticle-num} 
\bibliography{main}

\end{document}